\documentclass[final,3p,times]{article}
\usepackage{graphicx}
\usepackage{amssymb}
\usepackage[]{amsmath}
\usepackage{amsfonts}
\usepackage{float}
\def\PT{$\cal{PT}$}
\def\[{\begin{equation}}
\def\]{\end{equation}}
\def\PT{$\cal{PT}$}
\def\[{\begin{equation}}
\def\]{\end{equation}}
\usepackage{times}
\usepackage{anysize}
\marginsize{2.5cm}{2.5cm}{1cm}{2cm}
\selectfont

\begin{document}

\title{Dynamics of high-order solitons in the  nonlocal  nonlinear Schr\"{o}dinger equations}

\author{Bo Yang, Yong Chen\thanks{Address for
correspondence: Y.Chen, Shanghai Key Laboratory of Trustworthy Computing, East China Normal University, Shanghai, 200062 People's Republic of China, e-mail: ychen@sei.ecnu.edu.cn. }}

\maketitle

\begin{abstract}
A study of high-order solitons in three  nonlocal nonlinear Schr\"{o}dinger equations  is presented, which includes  the \PT-symmetric, reverse-time, and reverse-space-time nonlocal nonlinear Schr\"{o}dinger equations. General high-order solitons in three different  equations are derived from the same Riemann-Hilbert solutions of the AKNS hierarchy,  except for the difference in the  corresponding  symmetry relations on the ``perturbed" scattering data.  Dynamics of general high-order solitons in these equations is further analyzed.   It is shown that the high-order fundamental-soliton  is always moving on several different  trajectories in nearly equal velocities,  and they can be nonsingular or repeatedly collapsing, depending on the choices of the parameters.    It is also shown that the  high-order multi-solitons  could have  more complicated wave structures  and  behave  very differently from high-order fundamental solitons.  More interesting is the  high-order hybrid-pattern solitons, which are derived from combination of different size of block matrix in the Riemann-Hilbert solutions and thus they can describe a nonlinear interaction between several types of solitons.
\end{abstract}

\section{\label{sec:level1}Introduction}
As an significant subject in many branches of nonlinear science,  the integrable nonlinear wave equations and soliton theory has been studied for  many years \cite{Ablowitz1981,Zakharov1984,Faddeev1987,Ablowitz1991,Yang2010}.   Most of the integrable equations are local equations,  i.e.,  the solution¡¯s  evolution depends only  on the local solution value with its local space and time derivatives. Recently, a number of nonlocal integrable equations  were found  and triggered renewed interest in integrable systems.  The first such nonlocal equation was the  \PT-symmetric nonlinear Schr\"{o}dinger (NLS) equation\cite{AblowitzMussPRL2013,AblowitzMussNonli2016}:
\[ \label{e:PTNLS}
\textrm{i} q_t(x,t)=q_{xx}(x,t)+2 q^2(x,t)q^*(-x,t),
\]
where asterisk $*$  represents complex conjugation.  For this equation,  the evolution of the solution at location $x$  depends both on the local position $x$  and the distant nonlocal position $-x$.  This implies that the states of the solution at distant opposite locations are directly related,  reminiscent of quantum entanglement in pairs of particles.  This nonlocal integrable equation is distinctly different from local equations,   which makes it  mathematically
interesting. In the view of potential applications,  this  equation was linked to an unconventional system of magnetics\cite{Gadzhimuradov2016}. In addition, since  equation (\ref{e:PTNLS}) is parity-time (\PT ) symmetric,  it is  related to the concept of \PT-symmetry,  which is a hot research area of contemporary physics\cite{Yangreview}.

Nonlocal equation (\ref{e:PTNLS}) was actively investigated\cite{AblowitzMussPRL2013,AblowitzMussNonli2016,AblowitzMussSAPM2016,Gerdjikov2017,Ablowitz_arxiv,WenYan2016,HuangLing2017,Lakshmanan2017,BYJY2017,FengLuo2017,ShepelskyRHPLT2017,JYRTNLS2017,ChenZhang2018}.  Meanwhile,  many other nonlocal nonlinear integrable equations were also introduced  and studied with different space and/or time coupling\cite{AblowitzMussSAPM2016,AblowitzMussPRE2014,YanZY2015,Khara2015,ZhuRTSasa2017,BoTransformations,Fokas2016,Lou1,ZhoudNLS,HePPTDS,BYnonlocalDS,Zhu1,Zhu2,Gurses2017}.
Indeed, solution properties in several  nonlocal equations had been analyzed by the inverse scattering transform method, Darboux transformation or the bilinear method.  These new systems could reproduce  solution patterns  which  had  already been discovered in their local counterparts.  Moreover, interesting behaviors such as blowing-up(i.e., collapsing) solutions\cite{AblowitzMussPRL2013,BYJY2017,BoTransformations} and the  existence of novel richer structures were also revealed\cite{FengLuo2017,ZhuRTSasa2017,HePPTDS,BYnonlocalDS,Zhu1}.  A connection between nonlocal and local equations was discovered in~\cite{BoTransformations}, where it was shown that many nonlocal equations could be converted to local equations through  transformations.

In this article, we study  high-order solitons and their dynamics  in the \PT-symmetric NLS equation (\ref{e:PTNLS}) as well as  the
reverse-time NLS equation\cite{AblowitzMussSAPM2016}:
\[ \label{e:RTNLS}
\textrm{i}q_t(x,t)=q_{xx}(x,t)+2 q^2(x,t) q(x,-t),
\]
and  the reverse-space-time NLS equation\cite{AblowitzMussSAPM2016}:
\[ \label{e:RSTNLS}
\textrm{i}q_t(x,t)=q_{xx}(x,t)+2  q^2(x,t) q(-x,-t).
\]
Introducing the following  coupled Schr\"{o}dinger equations\cite{Ablowitz1981,Zakharov1984,Yang2010}:
\begin{eqnarray}
&& \textmd{i}q_t=q_{xx}-2q^2r, \label{qequation} \\
&& \textmd{i}r_t=-r_{xx}+2r^2q. \label{requation}
\end{eqnarray}
Then,  equations (\ref{e:PTNLS})-(\ref{e:RSTNLS}) can be respectively obtained from the coupled system (\ref{qequation})-(\ref{requation}) under nonlocal reductions
\begin{eqnarray}
&& r(x,t)= -q^*(-x,t), \label{qPTS}\\
&& r(x,t)= -q(x,-t), \label{qRTS}  \\
&& r(x,t)= -q(-x,-t).  \label{qRSTS}
\end{eqnarray}

As we know,  the inverse scattering transform method indicates that it is the poles of reflection coefficient (or zeros of the Riemann-Hilbert problem) that give rise to the soliton solutions. In \cite{JYRTNLS2017}, general N-solitons, which corresponds to N-simple poles in the spectral plane,  are derived for nonlocal equations (\ref{e:PTNLS})-(\ref{e:RSTNLS})  using the inverse scattering and Riemann-Hilbert method.  From this Riemann-Hilbert framework,   new types of multi-solitons with novel eigenvalue configurations in the spectral plane are discovered.   Therefore, as a more general case,  soliton solutions correspond to multiple poles, that is, the high-order solitons can be taken into consideration for nonlocal NLS equations  (\ref{e:PTNLS})-(\ref{e:RSTNLS}).

This kind of soliton have wide applications,  it  can describe a weak bound state of solitons  and may appear in the study of train propagation of solitons with nearly equal velocities and amplitudes but having a particular chirp\cite{Gagnon1994}. High-order soliton for several local equations, such as  the Sine-Gordon, nonlinear Schr\"{o}dinger, Kadomtsev-Petviashvili I  and Landau-Lifshitz equations,  have been investigated in several literature before\cite{Gagnon1994,WadatiIST1984,Villarroel1999,AblowitzKP2000,ShchesnovichYang2003,LingLL2015}.  To the best of our knowledge,  high-order soliton for the  nonlocal NLS equations (\ref{e:PTNLS})-(\ref{e:RSTNLS}) have never been reported.

In this article,  we derive the general high-order solitons in the \PT-symmetric, reverse-time, and reverse-space-time nonlocal NLS equations (\ref{e:PTNLS})-(\ref{e:RSTNLS}).
These high-order solitons are reduced  from the same Riemann-Hilbert solutions of the AKNS hierarchy with different symmetry relations on the ``perturbed" scattering data,  which  consist of the ``perturbed" eigenvalues as well as the corresponding   eigenfunctions.  Dynamics of these solitons are also explored.  We show that a generic feature for  high-order solitons in all the three nonlocal equations  is repeated collapsing,  resemble those in the (first-order) $N$-solitons for the nonlocal NLS equations (\ref{e:PTNLS})-(\ref{e:RSTNLS}). We also show that the high-order fundamental-soliton  describes  several  travelling waves moving on different trajectories with nearly equal velocities.  While the  high-order multi-solitons  could  have  more complicated wave and trajectory structures which behave very  differently from  the high-order fundamental-soliton.  For this pattern,  the corresponding  eigenvalue configuration always have equal  numbers of zeros with equal order in the upper and lower complex planes.  Moreover,  we find the high-order hybrid-pattern solitons,  which corresponds to novel  eigenvalue configurations, i.e., combinations between  zeros of unequal order in the upper and lower complex planes. These new patterns  can describe the  nonlinear interaction between several types of solitons,  and exhibit distinctively dynamical patterns which have not been found before.
\section{High-order solitons for general coupled Schr\"{o}dinger equations}
To derive  high-order solitons in equations  (\ref{e:PTNLS})-(\ref{e:RSTNLS}),  we need to start with the Riemann-Hilbert solutions of high-order solitons for the coupled Schr\"{o}dinger equations for given scattering data.  Then, imposing appropriate symmetry relations on the scattering data,  the high-order solitons for each nonlocal equations can be obtained.

The coupled system (\ref{qequation})-(\ref{requation})  admits the following Lax-pair\cite{Ablowitz1981,Zakharov1984}:
\begin{eqnarray}
 &&Y_{x}= -\textmd{i} \zeta \Lambda Y  + Q  Y,  \label{Laxpairxp}\\
 &&Y_{t}= 2\textmd{i} \zeta^2 \Lambda Y - 2\zeta  Q Y - \textmd{i} \Lambda \left( Q_{x}-Q^2 \right) Y,  \label{Laxpairtp}
\end{eqnarray}
where,
\begin{equation}
\Lambda=\textmd{diag}(1,-1),\
Q(x,t)=\left(
                                         \begin{array}{cc}
                                           0 & q(x,t) \\
                                           r(x,t) & 0 \\
                                         \end{array}
                                       \right).
\end{equation}
For localized functions $q(x,t)$  and $r(x,t)$,  the inverse scattering transform and the modern Riemann-Hilbert method was developed in\cite{Zakharov1984,ZakharovShabat1972,AKNSIST1974,ZakharovShabat1979}.  Following this Riemann-Hilbert treatment,  N-solitons  in coupled Schr\"{o}dinger system can be written as ratios of determinants~\cite{Faddeev1987,Yang2010} :
\begin{eqnarray} \label{N-solitonDetqr}
&& q(x,t)=2\textmd{i}\frac{\left|
\begin{array}{cc}
 M &  \overline{Y}_{2}^T \\
 Y_{1} & 0  \\
\end{array}
\right|}
{\left| M \right|}, \ \ \ r(x,t)=-2\textmd{i}\frac{\left|
\begin{array}{cc}
 M &  \overline{Y}_{1}^T \\
 Y_{2} & 0 \\
\end{array}
\right|}
{\left| M \right|},
\end{eqnarray}
where, $ Y=\left(v_{1}(x,t),...,v_{N}(x,t)\right), \overline{Y}=\left(\bar{v}_{1}(x,t),...,\bar{v}_{N}(x,t)\right)$.  $Y_{k}$  and $\overline{Y}_{k}$  represents the $k$-th row of matrix $Y$ and $\overline{Y}$, respectively.

Here $v_{k}(x,t)$  and $\bar{v}_{k}(x,t)$ are both column vectors given by
\begin{eqnarray}
&& v_{k}(x,t)=\exp[-i\zeta_{k}\Lambda x + 2i\zeta_{k}^2 \Lambda t]v_{k0},\label{Eigenfunction} \\
&& \bar{v}_{k}(x,t)=\exp[i\bar{\zeta}_{k}\Lambda x - 2i\bar{\zeta}_{k}^2 \Lambda t]\bar{v}_{k0}.\label{AdEigenfunction}
\end{eqnarray}

$M$  is a $N \times N$ matrix defined as:
\begin{eqnarray}
M=\left(M^{(N)}_{j,k}\right)_{1\leq j,k \leq N},\  \  M^{(N)}_{j,k}=\frac{\bar{v}_{j}^T v_{k}}{\bar{\zeta}_{j}-\zeta_{k}}, \ 1\leq j,k \leq  N,
\end{eqnarray}
here $\zeta_{k} \in \mathbb{C}_{+}$ (upper half complex  plane),  $\bar{\zeta}_{k} \in \mathbb{C}_{-}$ (lower half complex plane),  $v_{k0}$, $\bar{v}_{k0}$ are constant column vectors of length two.

With this formula,  the general high-order solitons can be directly obtained through a simple limiting process.  For this purpose, setting $N$ discrete spectral in the eigenfunction (\ref{Eigenfunction})  to be:
\begin{eqnarray*}
&&  \zeta_{2}=\zeta_{1}+\epsilon_{1,1}, \ldots,  \zeta_{n_{1}}=\zeta_{1}+\epsilon_{1,n_{1}-1},\\
&&   \zeta_{n_{1}+1}=\zeta_{2},\  \zeta_{n_{1}+2}=\zeta_{2}+\epsilon_{2,1},\ldots, \ \zeta_{n_{1}+n_{2}}=\zeta_{2}+\epsilon_{2,n_{2}-1},\\
&&  \cdots \\
&&   \zeta_{N-n_{r}+1}=\zeta_{r}, \ \zeta_{N-n_{r}+2}=\zeta_{r}+\epsilon_{r,1},\ldots,\  \zeta_{N}=\zeta_{r}+\epsilon_{r,n_{r}-1}.
\end{eqnarray*}
Similarly, setting another $N$ discrete  spectral in the adjoint eigenfunction (\ref{AdEigenfunction}) to be:
\begin{eqnarray*}
&&  \bar{\zeta}_{2}=\bar{\zeta}_{1}+\bar{\epsilon}_{1,1}, \ldots,  \bar{\zeta}_{\bar{n}_{1}}=\bar{\zeta}_{1}+\bar{\epsilon}_{1,\bar{n}_{1}-1},\\
&&  \bar{\zeta}_{\bar{n}_{1}+1}=\bar{\zeta}_{2}, \ \bar{\zeta}_{\bar{n}_{1}+2}=\bar{\zeta}_{2}+\bar{\epsilon}_{2,1},\ldots,\ \bar{\zeta}_{\bar{n}_{1}+\bar{n}_{2}}=\bar{\zeta}_{1}+\bar{\epsilon}_{2,\bar{n}_{2}-1},\\
&&  \cdots \\
&&  \bar{\zeta}_{N-\bar{n}_{s}+1}=\bar{\zeta}_{s},\  \bar{\zeta}_{N-\bar{n}_{1}+2}=\bar{\zeta}_{s}+\bar{\epsilon}_{s,1},\ldots, \ \bar{\zeta}_{N}=\bar{\zeta}_{s}+\bar{\epsilon}_{s,\bar{n}_{s}-1}.
\end{eqnarray*}
Here,   we should have $\sum_{i=1}^{r}n_{i}=\sum_{i=1}^{s}\bar{n}_{i}=N,$  and $r,s \in \mathbb{Z}_{+}$.

Then we have the following expansions:
\begin{eqnarray*}
&& v_{j}(\zeta_{j}+\epsilon_{j, k_{j}})=\sum_{k=0}^{\infty} v_{j}^{(k)} \epsilon_{j, k_{j}}^k,   \ \ \   \bar{v}_{i}(\bar{\zeta}_{i}+\bar{\epsilon}_{i, k_{i}})=\sum_{k=0}^{\infty} \bar{v}_{i}^{(k)} \bar{\epsilon}_{i, k_{i}}^k,  \\
 &&  \frac{\bar{v}_{i}^T(\bar{\zeta}_{i}+\bar{\epsilon}_{i, k_{i}}) v_{j}(\zeta_{j}+\epsilon_{j, k_{j}})}{\bar{\zeta}_{i}-\zeta_{j}+\bar{\epsilon}_{i, k_{i}}-\epsilon_{j, k_{j}}}
=\sum_{l=0}^{\infty}\sum_{k=0}^{\infty} M^{[k,l]}_{i,j}
\bar{\epsilon}_{i, k_{i}}^{\ l} \epsilon_{j, k_{j}}^k.
\end{eqnarray*}
Therefore, applying these expansions to each matrix element in N-soliton formula (\ref{N-solitonDetqr}), performing simple determinant manipulations and taking the limits of $\epsilon_{j, k_{j}},  \bar{\epsilon}_{i, k_{i}} \rightarrow 0 \  (k_{j}=1,..., n_{j}-1,\ k_{i}=1,..., \bar{n}_{i}-1)$, we derive the general high-order solitons for coupled Schr\"{o}dinger equations (\ref{qequation})-(\ref{requation}),  which are summarized in the following theorem.

\textbf{Theorem 1}.  \emph{The general high-order solitons in the coupled Schr\"{o}dinger equations (\ref{qequation})-(\ref{requation}) can be formulated as:}
\[ \label{highsoliton}
q(x,t)= 2 \textmd{i} \frac{\tau_{12}}{\tau_0},\ \ r(x,t)= -2 \textmd{i} \frac{\tau_{21}}{\tau_0},
\]
\emph{where}
\begin{eqnarray*}
&& \tau_0= \det\left( M\right),\ \ \tau _{kj}=\det\left(   \begin{array}{cc}
                       M & \bar{\phi}_{j}^{T} \\
                        \phi_{k} & 0
                      \end{array}
\right),\\
&& M=\left(M_{i,j}\right)^{{1\leq i \leq s}}_{1\leq j \leq r}, \ M_{i,j}=\left(M^{[k,l]}_{i,j}\right)^{0\leq k \leq \bar{n}_{i}-1}_{0\leq l \leq n_{j}-1,},
\end{eqnarray*}
\emph{and}
\begin{eqnarray*}
&&\phi=\left[\  v_{1}^{(0)} ,\ldots , v_{1}^{(n_{1}-1)},\ldots, v_{r}^{(0)},\ldots , v_{r}^{(n_{r}-1)} \right], \\
&&\bar{\phi}=\left[\  \bar{v}_{1}^{(0)}, \ldots , \bar{v}_{1}^{(\bar{n}_{1}-1)},\ldots,  \bar{v}_{s}^{(0)},  \ldots , \bar{v}_{r}^{(\bar{n}_{s}-1)}   \right].
\end{eqnarray*}
\emph{Here, $\phi_{k}$ stands the $k$-th row in matrix $\phi$,   so is for $\bar{\phi}_{j}$}.

This general soliton formula (\ref{highsoliton}) has been reported  in \cite{ShchesnovichYang2003} (Via using dressing method) as well as in \cite{LingLL2015} (By the generalized Darboux transformation).  So the proof of this theorem can be given along the lines of \cite{ShchesnovichYang2003,LingLL2015}.

\section{Symmetry relations of ``perturbed" scattering data in the nonlocal NLS equations}
We first recall  revelent results on symmetry relations of scattering data for the nonlocal NLS equations (\ref{e:PTNLS})-(\ref{e:RSTNLS}) presented in \cite{JYRTNLS2017}.   For this purpose, we denote:
\[
v_{k0}=\left[ a_{k}, b_{k} \right]^T,\ \
\bar{v}_{k0}=\left[\bar{a}_{k}, \bar{b}_{k} \right]^T.
\]
Next, with initial condition on the potential matrix:
\begin{equation}
Q_{0}:=Q(x)=\left(
                                         \begin{array}{cc}
                                           0 & q(x,0) \\
                                           r(x,0) & 0 \\
                                         \end{array}
                                       \right),
\end{equation}
here $q(x,0)$, $r(x,0)$  are the initial value of functions $q(x,t)$  and $r(x,t)$  at $t=0$.

Considering    the eigenvalue problem
\begin{equation}
  Y_{x}= -\textmd{i} \zeta \Lambda Y  + Q_{0}  Y, \label{eigen-problem}
\end{equation}
and its adjoint eigenvalue problem
\begin{equation}
  K^T_{x}= \textmd{i} \zeta K^T \Lambda -K^T Q_{0}. \label{adeigen-problem}
\end{equation}
Therefore, by using the symmetry of potential matrix $Q_{0}$  for each nonlocal reduction (\ref{qPTS})-(\ref{qRSTS}),  along with the large-$x$ asymptotics of $\zeta_{k}$'s eigenfunction $Y_{k}(x)$  as well as  $\bar{\zeta}_{k}$'s eigenfunction $K_{k}(x)$ , ref.\cite{JYRTNLS2017} derives the connections between each subset of scattering data  $\{\zeta_{k}, a_{k}, b_{k}\}$  and $\{\bar{\zeta}_{k}, \bar{a}_{k}, \bar{b}_{k}\}$ with rigorous proof.   Here, these important results can be directly used for our purpose.

In the following, we intend to show that:  through a  simple  modification to the  original scattering data,  new free  parameters can be introduced. In that case, we modify the existing  scattering data with a perturbation, i.e.,
\begin{eqnarray}\label{Perscattering1}
\{ \zeta_{k}, a_{k}, b_{k} \} \mapsto \{ \zeta_{k}(\epsilon), a_{k}(\epsilon), b_{k}(\epsilon) \},
\end{eqnarray}
where  $\zeta_{k}(\epsilon):= \zeta_{k} + \epsilon$,  and $a_{k}(\epsilon)$ and $b_{k}(\epsilon)$ can be further defined as:
\begin{eqnarray}\label{Perscattering2}
a_{k}(\epsilon):=e^{\phi_{0}+\phi_{1}\epsilon+\phi_{2}\epsilon^2+\cdots}, \ b_{k}(\epsilon):=e^{\varphi_{0}+\varphi_{1}\epsilon+\varphi_{2}\epsilon^2+\cdots}.
\end{eqnarray}
Here, $\phi_{k}$,  $\varphi_{j}$  are free complex parameters.

For the \PT-symmetric NLS equation (\ref{e:PTNLS}), following the derivation of \textbf{Theorem 1} in \cite{JYRTNLS2017}  and using the large-$x$ asymptotics of eigenfunctions, we can obtain the  symmetry relations of ``perturbed" scattering data (\ref{Perscattering1})-(\ref{Perscattering2}), which are summarized as: For a pair of non-imaginary eigenvalues  $(\zeta_{k},\ \hat{\zeta}_{k})\in \mathbb{C}_{+}$, $\hat{\zeta}_{k}= -\zeta^*_{k} $,  the corresponding  ``perturbed" eigenvalues are defined  as  $(\zeta_{k}(\epsilon),\ \hat{\zeta}_{k}(\epsilon))\in \mathbb{C}_{+}$,  where $\hat{\zeta}_{k}(\epsilon) \equiv -\zeta^*_{k}(\epsilon)$. After scaling the first element $a(\epsilon)$ to 1, the ``perturbed" eigenvectors $v_{k0}(\epsilon)$ and $\hat{v}_{k0}(\epsilon)$ are related as
\begin{eqnarray}\label{PTusymmetry1}
\hat{v}_{k0}(\epsilon)=\sigma_{1}v^*_{k0}(\epsilon),\ \ v_{k0}(\epsilon)=\left[1,  e^{\sum_{j=0}^{\infty}b_{k j}\epsilon^{j}}\right]^T, \ \ b_{k j}\in \mathbb{C}.
\end{eqnarray}
Repeating above arguments on the adjoint eigenvalue problem, we  have: for a pair of non-imaginary
$(\bar{\zeta}_{k},\ \hat{\bar{\zeta}}_{k})\in \mathbb{C}_{-}$, $\hat{\bar{\zeta}}_{k}= -\bar{\zeta}_{k}^*$,  the ``perturbed" eigenvalues are defined as  $(\bar{\zeta}_{k}(\bar{\epsilon}),\ \hat{\bar{\zeta}}_{k}(\bar{\epsilon}))\in \mathbb{C}_{-}$,  where $\hat{\bar{\zeta}}_{k}(\bar{\epsilon}) \equiv -\bar{\zeta}_{k}^*(\bar{\epsilon})$, and the form of their eigenvectors can be similarly obtained as
\begin{eqnarray}\label{AdPTusymmetry1}
\hat{\bar{v}}_{k0}(\bar{\epsilon})=\sigma_{1}\bar{v}^*_{k0}(\bar{\epsilon}),\
\bar{v}_{k0}(\bar{\epsilon})=\left[1,  e^{\sum_{j=0}^{\infty}\bar{b}_{kj}\bar{\epsilon}^{j}}\right]^T,\ \ \ \bar{b}_{kj}\in \mathbb{C}.
\end{eqnarray}

Especially, if $\zeta_{k}(\epsilon)$ is purely imaginary,   from above definition of ``perturbed" eigenvalues,  we have $\hat{\zeta}_{k}(\epsilon)=\zeta_{k}(\epsilon)$.  Because $-\zeta_{k}^*=\zeta_{k}$, thus, we have  $\epsilon^*=-\epsilon$.  In this case,  their ``perturbed" eigenvectors are also the same, which can be further scaled into the following form:
\begin{eqnarray}\label{Pertusymmetry1.1}
\hat{v}_{k0}(\epsilon)=v_{k0}(\epsilon)=\left[1,  e^{\sum_{j=0}^{\infty}(\textmd{i})^{j+1}\theta_{kj}\epsilon^{j}}\right]^T, \ \theta_{kj} \in \mathbb{R}.
\end{eqnarray}
Similarly, when $\bar{\zeta}_{k}(\epsilon)$ is also purely imaginary, its eigenvector is of the form:
\begin{eqnarray}\label{Pertusymmetry1.2}
\hat{\bar{v}}_{k0}(\bar{\epsilon})=\bar{v}_{k0}(\bar{\epsilon})=\left[1,  e^{\sum_{j=0}^{\infty}(\textmd{i})^{j+1}\bar{\theta}_{kj}\bar{\epsilon}^{j}}\right]^T, \ \bar{\theta}_{kj} \in \mathbb{R}.
\end{eqnarray}

Next,  for the reverse-time NLS equation (\ref{e:RTNLS}).  Following  the derivation of \textbf{Theorem 2} in \cite{JYRTNLS2017} as well as the above analysis,  we can also derive the symmetry relations of its ``perturbed" scattering data, which is represented as:  For a pair of discrete  eigenvalues $(\zeta_{k},  \bar{\zeta}_{k})$, where $\zeta_{k}  \in \mathbb{C}_{+}$  and $\bar{\zeta}_{k}=-\zeta_{k} \in \mathbb{C}_{-}$.  The  ``perturbed" eigenvalues are defined  as  $(\zeta_{k}(\epsilon),\ \bar{\zeta}_{k}(\bar{\epsilon}))$,  where $\zeta_{k}(\epsilon) \in \mathbb{C}_{+}$,  and  $\bar{\zeta}_{k}(\bar{\epsilon})\equiv -\zeta_{k}(-\bar{\epsilon})\in \mathbb{C}_{-}$. Then we scale the ``perturbed" eigenvectors $v_{k0}(\epsilon)$ and $\bar{v}_{k0}(\bar{\epsilon})$ with  their first elements become 1, and they are related as:
\begin{eqnarray}\label{RTSymmetryEigenvetors}
v_{k0}(\epsilon)=[1, e^{\sum_{j=0}^{\infty}b_{k,j}\epsilon^{j}} ]^{T},\  \bar{v}_{k0}=v_{k0}(-\bar{\epsilon}),\ \ b_{kj}\in \mathbb{C}.
\end{eqnarray}
where $b_{k}$  is an arbitrary complex parameter.

However, for the reverse-space-time NLS equation (\ref{e:RSTNLS}),  according to the  symmetry relations  on the scattering data given by \textbf{Theorem 3} in \cite{JYRTNLS2017}, i.e.,  the  eigenvalues  $\zeta_{k}$  can be anywhere in $\mathbb{C}_{+}$  and  $\bar{\zeta}_{k}$  can be anywhere in $\mathbb{C}_{-}$.  And  the corresponding eigenvectors must be of the forms
\begin{eqnarray}\label{RSTSymmetryEigenvetors}
  v_{k0}=[1,\omega_{k}]^{T},\ \omega_{k}=\pm 1;  \ \ \   \bar{v}_{k0}=[1,\bar{\omega}_{k}]^{T},\ \bar{\omega}_{k}=\pm 1.
\end{eqnarray}
We find that all the parameters in the ``perturbed" scattering data can be eliminated in the ``perturbed" eigenvectors. Thus, no more parameters can be introduced in (\ref{RSTSymmetryEigenvetors}) so that  we have  $v_{k0}(\epsilon)=v_{k0}, \ \bar{v}_{k0}(\bar{\epsilon})=\bar{v}_{k0}$.

Therefore, utilizing the above symmetry  relations  on the ``perturbed" scattering data in the high-order Riemann-Hilbert solution (\ref{highsoliton}),  we will  construct high-order solitons for  nonlocal NLS equations (\ref{e:PTNLS})-(\ref{e:RSTNLS}) in the sections below.

\section{Dynamics of high-order solitons in the  $\mathcal{PT}$-symmetric nonlocal  NLS equation}
To derive the $N$-th order solitons in the $\mathcal{PT}$-symmetric  NLS equation  (\ref{e:PTNLS}),  we just need to apply corresponding symmetry  relations of the scattering data  to the general soliton formula (\ref{highsoliton}). Then we investigate solution dynamics in the high-order fundamental (one)-soliton as well as the high-order multi-solitons.

\subsection{High-order fundamental-soliton}
Firstly,  we consider the second-order fundamental-soliton,  which corresponds to a single pair of purely imaginary eigenvalues (zero of multiplicity two)  $\zeta_{1}=i \eta_{1} \in i\mathbb{R}_{+}$,  and $\bar{\zeta}_{1}=i \bar{\eta}_{1} \in i\mathbb{R}_{-}$,  where $\eta_{1} >0$  and $\bar{\eta}_{1}<0$,  In this case, symmetry relations on the perturbed  eigenfunctions are given by (\ref{Pertusymmetry1.1})-(\ref{Pertusymmetry1.2}), i.e.,  $v_{10}(\epsilon)=\left[1, e^{i \theta_{10}-\theta_{11}\epsilon\ }\right]^T$,  and $\bar{v}_{10}(\bar{\epsilon})=\left[1, e^{i \bar{\theta}_{10}-\bar{\theta}_{11}\bar{\epsilon}\ }\right]^T$, where $\theta_{10}, \theta_{11}, \bar{\theta}_{10}, \bar{\theta}_{11}$  are real constants. Substituting these expressions into formula (\ref{highsoliton}) with $N=n_{1}=\bar{n}_{1}=2$, we obtain the  analytic expression for   the second-order fundamental soliton of equation (\ref{e:PTNLS}):
\begin{eqnarray}\label{RS2-ndHighSoliton}
q(x,t)=\frac{2(\bar{\eta}_{1}-\eta_{1}) \left[\mathcal{G}(x,t)e^{2 \bar{\eta}_{1} x-4i \bar{\eta}_{1}^2 t+i \bar{\theta}_{10}}+\bar{\mathcal{G}}(x,t)e^{2 \eta_{1} x-4i \eta_{1}^2 t-i \theta_{10}} \right]}{4\cosh^2\left[(\eta_{1}-\bar{\eta}_{1}) x-2i(\eta_{1}^2-\bar{\eta}_{1}^2)t-\frac{i}{2}(\theta_{10}+\bar{\theta}_{10})\right]+\mathcal{F}(x,t)}, \ \
\end{eqnarray}
where  $\mathcal{F}(x,t)=-(\mathcal{G}+2)(\bar{\mathcal{G}}+2)$,  with
\begin{eqnarray}
&&\mathcal{G}(x,t)=(\bar{\eta}_{1}-\eta_{1})(2x-8 i \eta_{1}t+i\theta_{11})-2,  \label{Npolynomialterms1}\\
&&\bar{\mathcal{G}}(x,t)=(\eta_{1}-\bar{\eta}_{1})(2x-8 i \bar{\eta}_{1}t-i \bar{\theta}_{11})-2. \label{Npolynomialterms2}
\end{eqnarray}
This kind of soliton,  which  combines  exponential functions with algebraic polynomials, has never been reported before in the nonlocal NLS equation (\ref{e:PTNLS}). It contains  six real parameters: $\eta_{1}, \bar{\eta}_{1}, \theta_{10}, \bar{\theta}_{10}, \theta_{11}$  and $\bar{\theta}_{11}$. The motion trajectory for this solution  can be approximatively described by the following two curves
\begin{eqnarray}
 \Sigma_{\pm}:\ \  2(\bar{\eta}_{1}-\eta_{1})x  \pm  \ln  \left|\mathcal{F}(x,t)\right|=0.  \ \  (\left|\mathcal{F}(x,t)\right| \neq 0) \label{RSAsymptoticline1}
\end{eqnarray}
In this case, two solitons moving along the center trajectories $\Sigma_{+}$  and $\Sigma_{-}$.  When $|x|\rightarrow \pm \infty$,  the amplitude  $|q|$ of  the  solution   decays  exponentially to zero.    However, with the development of time,   a simple asymptotic analysis with estimation on the leading-order terms  shows that: when  soliton (\ref{RS2-ndHighSoliton}) is moving on $\Sigma_{+}$  or $\Sigma_{-}$, its amplitudes $|q|$ can approximately vary as:
\begin{eqnarray}\label{EstimationPTNLS}
|q(x,t)|\sim \frac{2|\eta_{1}-\bar{\eta}_{1}|e^{\left(\eta_{1}+\bar{\eta}_{1}\right)z(x,t)}}{|e^{\pm 2i\gamma t -i\tau_{0} \pm i (\theta_{10}+\bar{\theta}_{10})}+1|},\ \ \  t \sim \pm \infty,
\end{eqnarray}
where  $z(x,t)=\frac{\ln|\mathcal{F}(x,t)|}{\pm 2\left(\eta_{1} - \bar{\eta}_{1}\right)}$,
$\gamma=2(\bar{\eta}_{1}^2-\eta_{1}^2)$, $\tau_{0}= \text{Arg} \left[\mathcal{F}(x,t)\right]+2 k \pi,\ (k \in \mathbb{Z})$,  the positive and negative sign in (\ref{EstimationPTNLS}) respectively corresponds to $\Sigma_{+}$  and  $\Sigma_{-}$. (It should be noted that estimation (\ref{EstimationPTNLS}) is valid only when $|t| \gg \max\{|\theta_{11}|, |\bar{\theta}_{11}|\}$.  Before this,  the amplitudes $|q|$ of solution are unequal when soliton moves on each curve, depending on the value of parameter $\theta_{11}$ and $\bar{\theta}_{11}$.)

In the case when  $\eta_{1}=-\bar{\eta}_{1}$,  solution (\ref{RS2-ndHighSoliton})  will be nonsingular or collapsing at certain locations,  depending on the values of these parameters. Specifically,

(1). If $\theta_{11}=\bar{\theta}_{11}$,  as long as $\theta_{10}+\bar{\theta}_{10} \neq(2k+1)\pi$ for any integer $k$, this soliton  will be nonsingular.

(2). If  $\theta_{11}\neq\bar{\theta}_{11}$,   we first define three  multivariate functions,  these are  $\emph{c}_{0}\equiv\frac{\sin(\theta_{10}+\bar{\theta}_{10})}{ \eta_{1}(\theta_{11}-\bar{\theta}_{11})}$,  $\Delta_{1}\equiv\frac{(\theta_{11}-\bar{\theta}_{11})^2}{4}-\frac{1+\cos(\theta_{10}-\bar{\theta}_{10})}{2\eta_{1}^2}$  and $\Delta_{2}\equiv -4x_{c}^2+\frac{(\theta_{11}-\bar{\theta}_{11})^2}{4}-\frac{1+\cosh(4\eta_{1}x_{c})\cos(\theta_{10}-\bar{\theta}_{10})}{2\eta_{1}^2}$, which contain all the parameters. Then, solution (\ref{RS2-ndHighSoliton}) will not blow up only when $c_{0}$,  $\Delta_{1}$  and $\Delta_{2}$  satisfy one of the following two critical conditions:
\begin{eqnarray}
&& \textrm{(a)}. \ \textmd{For any} \ \theta_{10}+\bar{\theta}_{10}\neq(2k+1)\pi,\ \emph{c}_{0} \notin (0,1), \  \textmd{and} \ \Delta_{1} <0.  \\
&& \textrm{(b)}. \ \emph{c}_{0} \in (0,1),   \  \textmd{and} \ \Delta_{1}, \Delta_{2}  <0.
\end{eqnarray}
Otherwise, when $\Delta_{1}\geq0$ in condition (a), there will be two(or one) singular points locating at $x=0$, $t=\frac{\pm \sqrt{\Delta_{1}}}{8\eta_{1}}+t_{0}$, where  $t_{0}=\frac{\theta_{11}-\bar{\theta}_{11}}{16\eta_{1}}$.  Or, when $\Delta_{2}\geq0$  with $c_{0}\in (0,1)$ in  (b), there would also have two(or one) singular points locating at $x=x_{c}$, $t=\frac{\pm \sqrt{\Delta_{2}}}{8\eta_{1}}+t_{0}$.  Here $x_{c}$  admits a special transcendental equation  $\frac{4\eta_{1}x_{c}}{\sinh(4\eta_{1}x_{c})}=c_{0}$,  which can be solved numerically for this given $c_{0}$.

Moreover, for all  the nonsingular solution, $|q(x,t)|$  reaches its peak amplitude at $x=0$, $t=t_{0}$ with the value attained as $\left|\frac{8\eta_{1}\left[\eta_{1}(\theta_{11}-\bar{\theta}_{11})\sin(\phi_{0})-2\cos(\phi_{0})\right]}{4\cos^2(\phi_{0})-\eta_{1}^2(\theta_{11}-\bar{\theta}_{11})^2}\right|$, where $\phi_{0}=\frac{\theta_{10}+\bar{\theta}_{10}}{2}$. When $t\rightarrow \pm \infty$,  according to a logarithmic law for large values of $|t|$, two solitons  moving along  $\Sigma_{+}$  and $\Sigma_{-}$ with almost equal velocities and amplitudes, and  peak amplitude does not exceed $\left|\frac{4\eta_{1}}{1+e^{i(\theta_{1}+\bar{\theta}_{1})}}\right|$.  To demonstrate, we choose the following parameters:
\begin{eqnarray}\label{ParameterRSNs}
\eta_{1}=0.5,\  \theta_{10}=\pi/4,\ \bar{\theta}_{10}=\pi/6,\  \theta_{11}=0.25,\ \bar{\theta}_{11}=0.5.
\end{eqnarray}
Propagation of this high-order soliton is displayed in Fig.1.
\begin{figure}[htbp]
   \begin{center}
     \vspace{-0.30cm}
     \hspace{-7.50cm} \includegraphics[scale=0.185, bb=0 0 385 567]{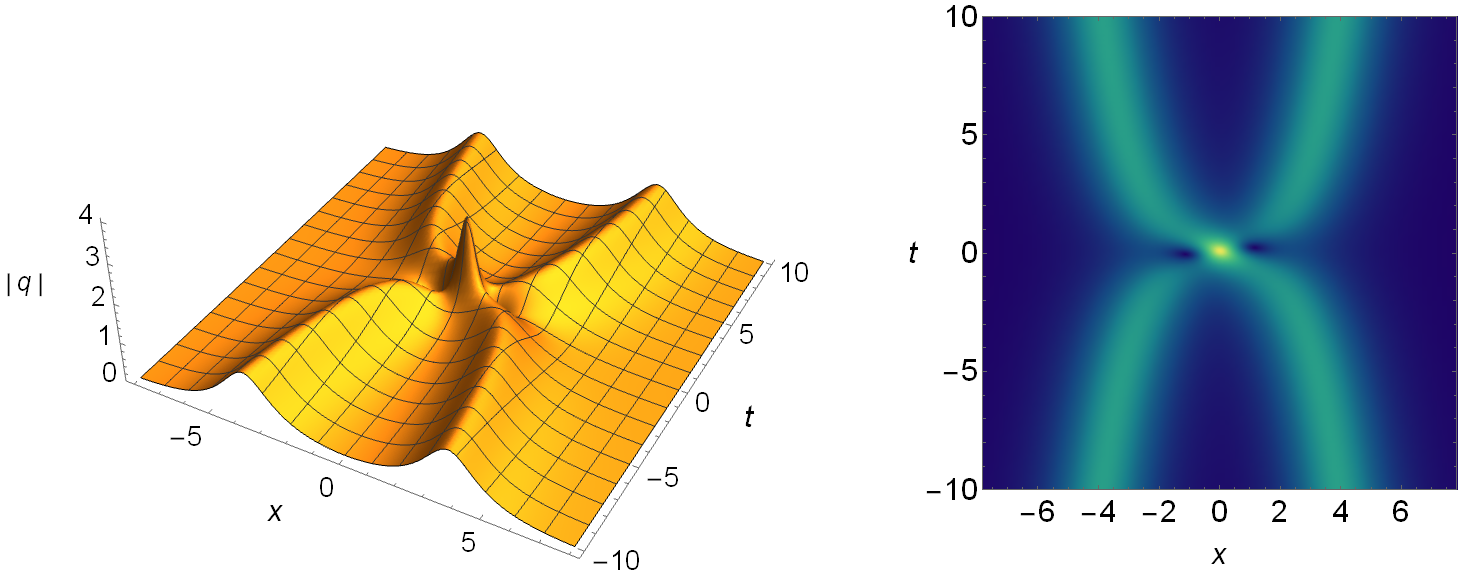}
   \caption{Left panel is the second-order one-soliton (\ref{RS2-ndHighSoliton}) with parameters (\ref{ParameterRSNs}). Right panel is the  corresponding density plot.}
   \label{Tu1}
     \hspace{-7.45cm} \includegraphics[scale=0.185, bb=0 0 385 567]{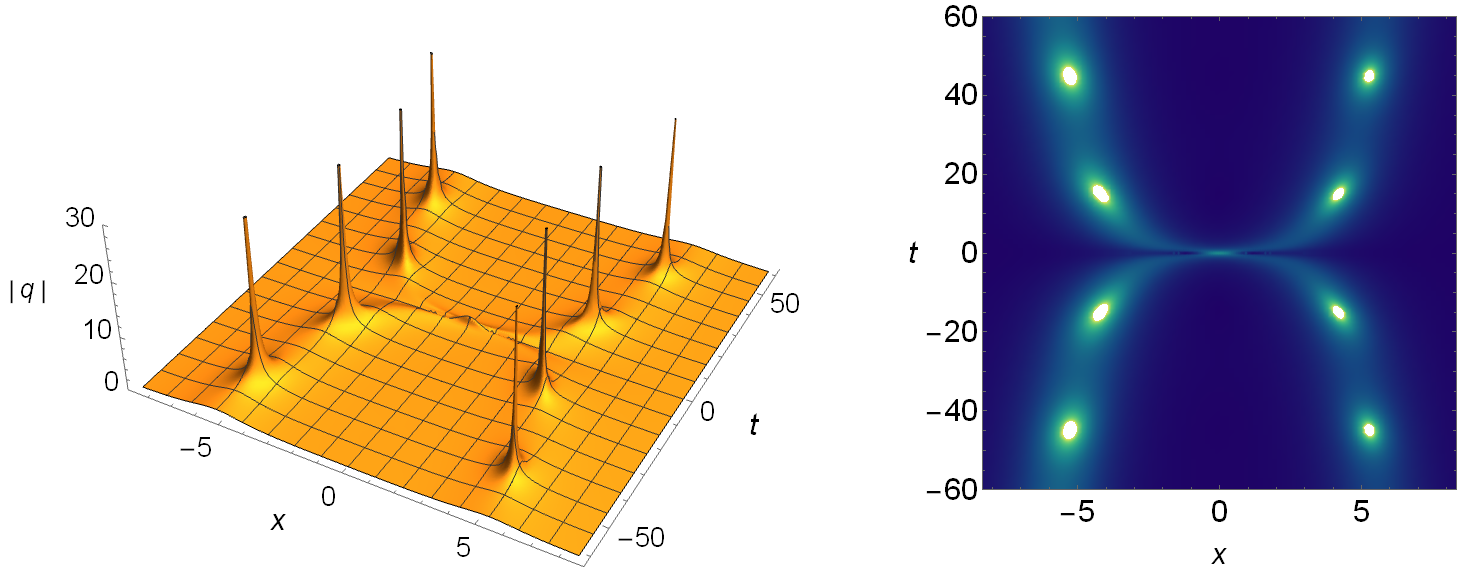}
   \caption{Left panel is the 2nd-order one-soliton (\ref{RS2-ndHighSoliton}) in eq.(\ref{e:PTNLS}) with parameters (\ref{ParameterRS1}).  Right panel  is the corresponding density plot.(Here, the bright spots shown on the density plot represent the location of singularity.)}
   \vspace{-0.75cm}
   \label{Tu2}
   \end{center}
\end{figure}

It is shown that two solitons are slowly moving in the  spatial orientation. This is quite different from  the  dynamics  of fundamental soliton in \cite{JYRTNLS2017},  where the soliton can not move in space.  The peak amplitude of $|q(x,t)|$ reaches about $2.65834$  at the location $(0,0.09375)$. Moreover,  with the evolution of time,  they keep almost identical value of maximum amplitudes, which is no larger than about 1.26047.

In a more general case,  where  $\eta_{1} \neq -\bar{\eta}_{1}$,  an important feather for this high-order soliton is the repeatedly collapsing  along two trajectories. This can be clarified from the large-time estimation  (\ref{EstimationPTNLS}).  Actually, when $|t|$ becomes very large, a direct calculation shows that $\lim_{|t|\rightarrow\infty}\text{Arg}\left[\mathcal{F}(x,t)\right]= \pi$. Thus, one can  repeatedly choose  large time point $t_{c}$ s.t.  $\cos(2\gamma t_{c} \mp\tau_{0}+(\theta_{10}+\bar{\theta}_{10}))=-1$. This implies the existence of singularities for the solution at large time.

Moreover, due to the impact of algebraic polynomial terms,  the  collapsing interval for this high-order soliton is no more a fixed value. Instead,  this so-called ``period " is slightly varying over time. Besides, amplitudes of solution $|q|$  are unequal when soliton moves on each path,  depending on the sign of $\eta_{1}+\bar{\eta}_{1}$.   To illustrate,  we choose parameters as
\begin{eqnarray}\label{ParameterRS1}
\eta_{1}=0.50,\ \ \bar{\eta}_{1}=-0.55,\ \ \theta_{10}=\bar{\theta}_{10}=0,\ \ \theta_{11}=\bar{\theta}_{11}=0.
\end{eqnarray}
Graphs of corresponding   second-order fundamental-soliton are shown in Fig.2.
Through simple  numerical calculation and approximate estimation,  the first singularities  quartet for this soliton  is  obtained,  which  locates approximately  at  $(\pm x_{c}, \pm t_{c})$ with  $x_{c} \approx 3.9999,\ t_{c} \approx 15.0169$,  and the first time interval between two successive singularities $\pm t_{c}$ is $30.0338$.  Afterwards,  the second singularities  quartet approximately appears  at  $(\pm \tilde{x}_{c}, \pm \tilde{t}_{c})$ with  $\tilde{x}_{c} \approx 5.0369,\ \tilde{t}_{c} \approx 44.9041$.  So the second  time interval between $\tilde{t}_{c}$  and $t_{c}$ is  about $29.9232$.

Generally,   the $N$-th order fundamental-soliton solution can be obtained in the same way by  choosing  $n_{1}=\bar{n}_{1}=N$ in formula (\ref{highsoliton}),  and the dynamics of $N$-wave motion on $N$ different asymptote trajectories  can be expected.
\subsection{High-order multi-solitons}
Now, we consider the  high-order   multi-solitons for the $\mathcal{PT}$-symmetric  NLS equation.  From the symmetries of  scattering data, the eigenvalues in the upper and lower halves of the complex plane are completely independent. This allows  for novel eigenvalue configurations, which gives rise to new types of high-order solitons with intersting dynamical patterns.   These results can be divided into the following two cases in principle:

\subsubsection{The normal pattern: Square-matrix blocks.}
For the  most normal pattern,  each  block  $\left(M^{[k,l]}_{i,j}\right)^{0\leq k \leq \bar{n}_{i}-1}_{0\leq l \leq n_{j}-1,}$  of  $\left(M_{i,j}\right)^{{1\leq i \leq s}}_{1\leq j \leq r}$ in formula(\ref{highsoliton}) is an square matrix.    In this case,  one has to  take the same index $s=r=m$  with $n_{k}=\bar{n}_{k}=n\ (k=1,2,..,m)$ and $N=n \times m$ in   (\ref{highsoliton}).  This yields the normal $N$-th order $m$-solitons.

For example,  we consider the second-order two-soliton.  Especially, choosing a pair of non-purely-imaginary eigenvalues: $\zeta_{1},\ \zeta_{2}  \in \{\mathbb{C}_{+} \setminus i\mathbb{R}_{+}\}$ with $\zeta_{2} =-\zeta_{1}^*$, which belongs to  the  second type two-solitons for eq.(\ref{e:PTNLS}) discussed   in \cite{JYRTNLS2017}.  Thus, from above results (\ref{PTusymmetry1})-(\ref{AdPTusymmetry1}),  their perturbed eigenvalues and  eigenvectors are related as
\begin{eqnarray*}
\zeta_{2}(\epsilon)= -\zeta^*_{1}(\epsilon),\ \  v_{20}(\epsilon)=\sigma_{1}v^*_{10}(\epsilon),\ v_{10}(\epsilon)=[1, e^{b_{10}+b_{11}\epsilon\ }]^T,
\end{eqnarray*}
where $b_{10}, b_{11}$  are complex constants.

Similarly, for  a pair of non-purely-imaginary eigenvalues $\bar{\zeta}_{1},\ \bar{\zeta}_{2}  \in \{\mathbb{C}_{-} \setminus i\mathbb{R}_{-}\}$,  with $\bar{\zeta}_{2} =-\bar{\zeta}_{1}^*$, their perturbed eigenvalues and  eigenfunctions are related as
\begin{eqnarray*}
 \bar{\zeta}_{2}(\bar{\epsilon})=-\bar{\zeta}_{1}^*(\bar{\epsilon}),\ \ \bar{v}_{20}(\bar{\epsilon})=\sigma_{1} \bar{v}_{10}^*(\bar{\epsilon}),\ \bar{v}_{10}(\epsilon)=[1, e^{\bar{b}_{10}+\bar{b}_{11}\epsilon\ }]^T,
\end{eqnarray*}
where $\bar{b}_{10}, \bar{b}_{11}$  are complex constants.
Substituting these data into (\ref{highsoliton}) with $N=4$, $n_{1}=n_{2}=2$ and $\bar{n}_{1}=\bar{n}_{2}=2$.   Then  it is found the corresponding aolution can be nonsingular or repeatedly collapse in pairs at spatial locations.   In addition,  they  can move in four opposite directions  and exhibit more complex wave-front structures.
\begin{figure}[htbp]
   \begin{center}
   \vspace{-1.25cm}
     \hspace{-4.5cm} \includegraphics[scale=0.240, bb=0 0 385 567]{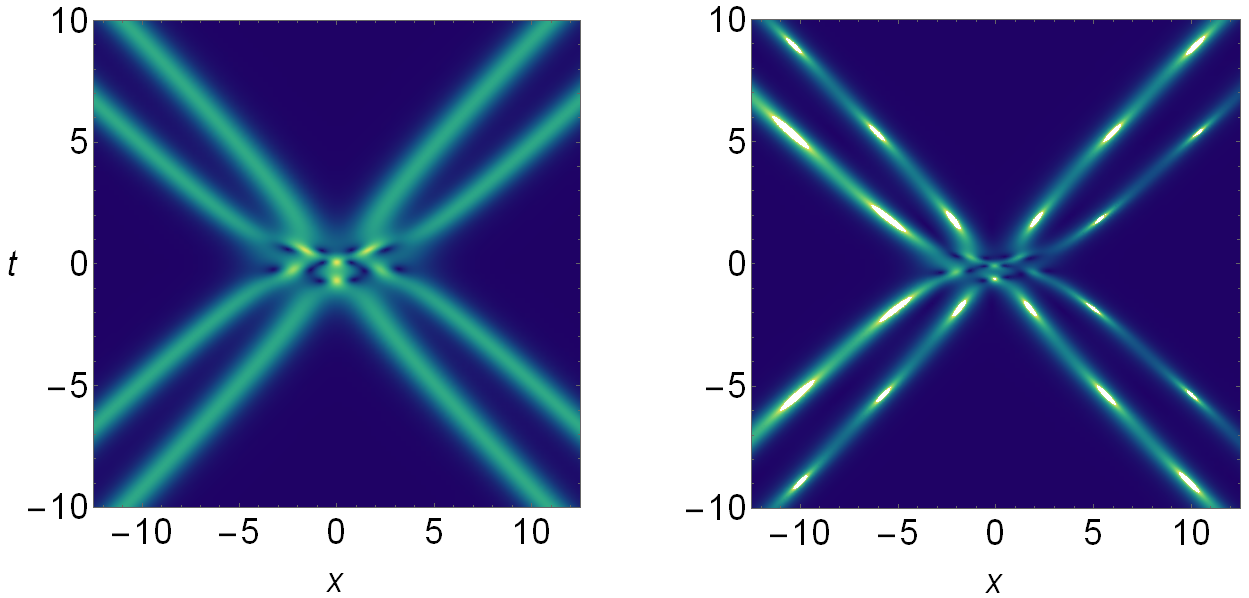}
   \caption{(a) is the  second-order two-solitons   with parameters (\ref{ParameterRS3-1})-(\ref{ParameterRS3-2}).
   (b) shows the  second-order two-solitons with parameters (\ref{ParameterRS4-1})-(\ref{ParameterRS4-2}).}
   \vspace{-0.55cm}
   \label{Tu1}
   \end{center}
\end{figure}

To demonstrate their dynamics, we choose two sets  of parameters:
\begin{eqnarray}
&&  \zeta_{1}=-\zeta_{2}^*=0.3+0.8i,\  \bar{\zeta}_{1}=-\bar{\zeta}_{2}^*=0.3-0.8i,  \label{ParameterRS3-1} \\
&& e^{b_{10}}=1+ 0.2i, \ e^{\bar{b}_{10}}=1-0.1 i.  \ e^{b_{11}}=0.2,  \ e^{\bar{b}_{11}}=0.25.    \label{ParameterRS3-2} \\
&&   \zeta_{1}=-\zeta_{2}^*=0.3+i, \bar{\zeta}_{1}=-\bar{\zeta}_{2}^*=0.3 - 1.2i,  \label{ParameterRS4-1}\\
&&  e^{b_{10}}=1+ i, \ e^{\bar{b}_{10}}=1- i,\  e^{b_{11}}=1, \  e^{\bar{b}_{11}}=1. \label{ParameterRS4-2}
\end{eqnarray}
Parameter (\ref{ParameterRS3-1})-(\ref{ParameterRS3-2}) generates a nonsingular solution which is plotted in the left panel of Fig.3,  while the right panel in Fig.3 exhibits the blowing-up solution  derived from parameter set (\ref{ParameterRS4-1})-(\ref{ParameterRS4-2}).   Especially, if the real parts of eigenvalues $\zeta_{k}$ and $\bar{\zeta}_{k}$ are not equal,  the amplitudes of  moving waves decreases or increases exponentially with time.
\subsubsection{The hybrid pattern: Combination of different block types.}
Secondly,  we consider a more general case, where the blocks (sub-matrices) are not required to be square matrices. Instead,  different types of blocks can be combined together through formula (\ref{highsoliton}).  Specifically,  defining two index sets $I_{1}$  and $I_{2}$ for the block matrix:
$I_{1}=\{n_{1},...,n_{r} \},\ I_{2}=\{ \bar{n}_{1},...,\bar{n}_{s} \}.$  From above discussion we know that $I_{1}$  and $I_{2}$ are mutually independent.  By virtue of this fact, novel patterns of solitons can be   achieved by taking different index values.   These  interesting hybrid patterns  have  not been reported before and can   describe the  interactions between several one- or multi-solitons  with unequal  orders.

Taking $N=2$ in  formula (\ref{highsoliton}),  then index sets  have three kinds of combinations(Regardless of other equivalent cases):  (a). $I_{1}=I_{2}=\{1, 1\}$; (b). $I_{1}=I_{2}=\{2\}$;  (c).  $I_{1}=\{1,1\},\  I_{2}=\{2\}$.  The first two combinations  are the normal case,  which corresponding to the two-soliton and second-order fundamental-soliton.   For the last one,  two simple zeros  which are symmetric about the imaginary axis  locates in $\mathbb{C}_{+}$  and one zero (multiplicity two) locates in $\mathbb{C}_{-}$. This  interesting  configuration of eigenvalues  corresponds to a special ``two-soliton" solution.   Such an example is shown  in Fig.4 with parameters:
\begin{eqnarray}\label{ParameterRS5}
\zeta_{1}=-\zeta_{2}^*=0.1+0.5i,\ \ \bar{\zeta_{1}} = -0.25i,\ \ b_{10}=0,\ \  \bar{\theta}_{10}=0,\ \  \bar{\theta}_{11}=0.2.
\end{eqnarray}
This soliton describes two waves traveling in opposite directions as they repeatedly collapsing over time. Remarkably, their motion trajectory  is no longer  straight line but on certain curves, which is different from the normal two-soliton.  In addition,   the amplitudes $|q|$  of two travelling waves  are   growing  or  decreasing  exponentially with time,  just along the directions of  motion.

\begin{figure}[htbp]
   \begin{center}
   \vspace{-0.25cm}
     \hspace{-7.5cm} \includegraphics[scale=0.173, bb=0 0 385 567]{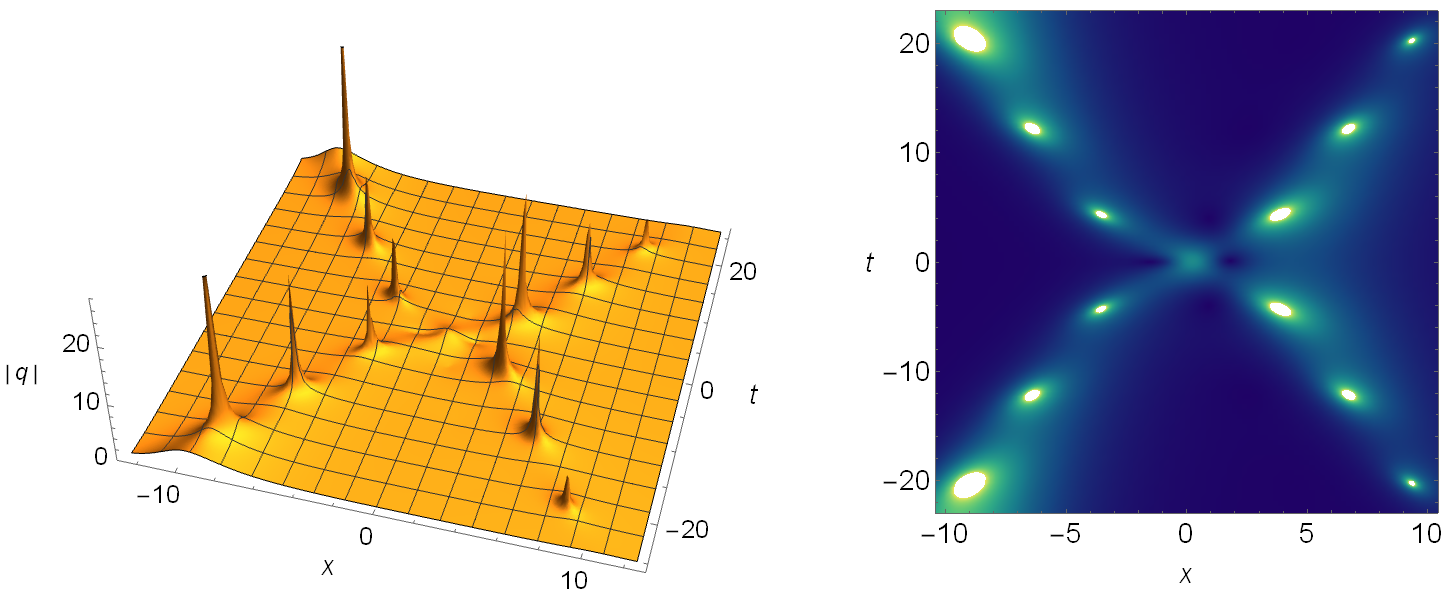}
   \caption{Left panel  is a hybrid solution with parameters (\ref{ParameterRS5}).\
   Right panel is the corresponding density plot. (Here, the   collapsing points are shown by the white bright spots,  when they are amplifying or shrinking along the line,  it means the solutions' amplitudes  are increasing or decreasing correspondingly.) }
    \vspace{-0.75cm}
   \label{Tu4}
   \end{center}
\end{figure}

Next, when $N=3$, the corresponding block sets have six combinations: (a). $I_{1}=I_{2}=\{1,1,1\}$; (b). $I_{1}=\{1,1,1\},\  I_{2}=\{1,2\}$;  (c).  $I_{1}=\{1,1,1\},\  I_{2}=\{3\}$;  (d). $I_{1}=\{1,2\},\ I_{2}=\{1,2\}$;  (e). $I_{1}=\{1,2\}, \ I_{2}=\{3\}$;  (f).  $I_{1}=I_{2}=\{3\}$.  These sets can feature the interactions of several types of one- or multi-solitons with certain orders,  except for the normal case (a)  and (f).

Specifically,  if we consider   combination $(b)$, there will be three simple pole in the upper half plane and one double-pole with one simple pole in the  lower half plane.  This eigenvalue configuration  can  also bring  new hybrid patterns,  which feathers  nonlinear superposition between a special ``two-soliton" and  a fundamental one-soliton.   Using parameter values
\begin{eqnarray}
&& \zeta_{1}=-\zeta_{2}^*=0.1+0.6i,\  \zeta_{3}=0.5i,\  \bar{\zeta_{1}}=-0.7i, \ \bar{\zeta_{2}}=-0.25i, \label{ParameterRS71}\\
&& b_{10}=0, \ \theta_{30}=\bar{\theta}_{10}=\bar{\theta}_{20}=\bar{\theta}_{21}=0.  \label{ParameterRS7}
\end{eqnarray}
The  associated solution is plotted in Fig.5.
This soliton feathers two waves travelling in two opposite curves, plus another stationary  wave (fundamental soliton) at $x=0$,  while they both collapse repeatedly along the  directions.  Moreover, the amplitudes of  the moving waves  are changing  with time as well.

Consider combination $(d)$ as another example.  In this case, there is one simple pole and one double-pole in the upper half plane  as well as  the  lower half plane.  This eigenvalue configuration  could create a new type of hybrid soliton which  differs from other patterns.   To illustrate its dynamics,  we  choose parameters
\begin{eqnarray}
&& \zeta_{1}=\bar{\zeta}_{1}^*=0.25i,\  \zeta_{2}=\bar{\zeta}_{2}^*=0.5i, \ \theta_{10}=-\bar{\theta}_{20}=-\pi/6, \label{ParameterRS6} \\
&& \bar{\theta}_{10}=-\theta_{20}=\pi/4,\ \theta _{21}=1, \ \bar{\theta}_{21}=0.5.
\end{eqnarray}
Corresponding graph for this solution  is presented in  Fig.5,  which feathers  the nonlinear interaction  between the  second-order one-soliton and a fundamental soliton.
\begin{figure}[htbp]
   \begin{center}
   \vspace{-0.25cm}
     \hspace{-7.5cm} \includegraphics[scale=0.175, bb=0 0 385 567]{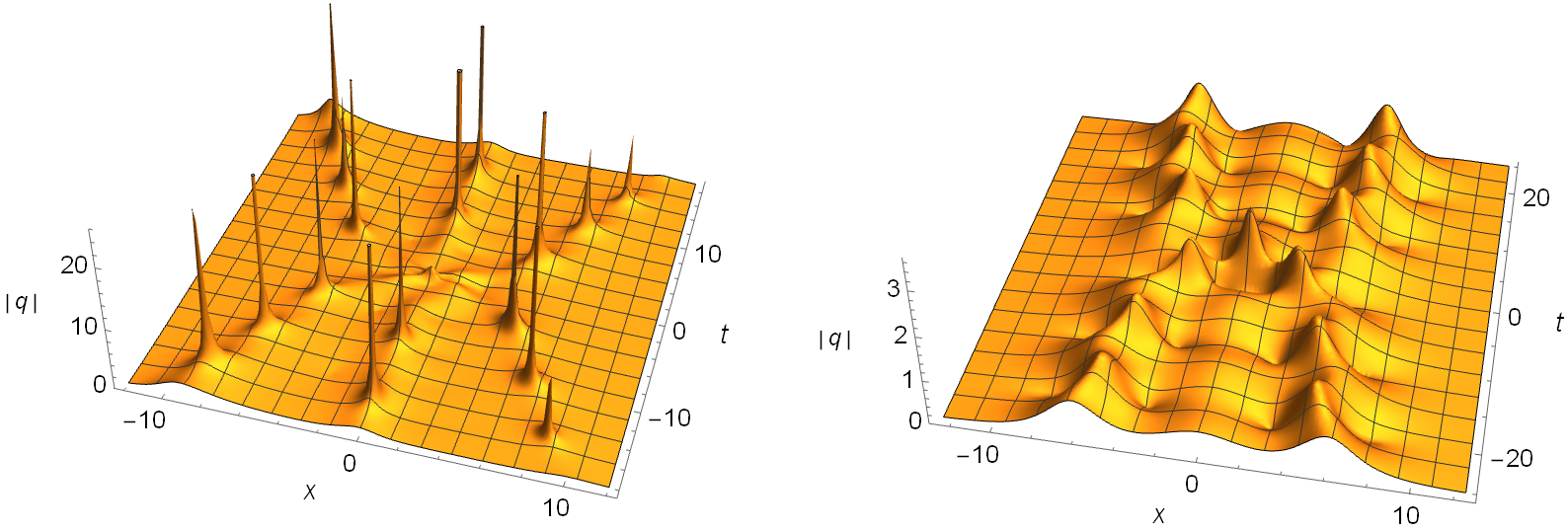}
   \caption{Left panel  is a hybrid solution with parameters (\ref{ParameterRS6}).
   Right panel shows a hybrid solution with parameters (\ref{ParameterRS71})-(\ref{ParameterRS7}). }
  \vspace{-0.75cm}
   \label{Tu1}
   \end{center}
\end{figure}

This soliton does not collapse  and the interesting  periodic phenomenon can be seen.  For the rest of the combinations,  we can still utilize formula (\ref{highsoliton})   to generate other  hybrid  patterns of solitons.

Therefore, as we could see,    the hybrid pattern solitons exhibit several new types of dynamics which have not been observed before. Similarly,  the higher order multi-hybrid solitons can be also investigated in this way and the additional novel phenomenon can be  expected.
\section{Dynamics of high-order solitons in the  reverse-time   NLS equation}
To derive  $N$-th order solitons for the reverse-time   NLS equation (\ref{e:RTNLS}), we need to impose corresponding symmetry  relations of ``perturbed" discrete  scattering data  in the general soliton formula (\ref{highsoliton}). Normally, for a pair of discrete eigenvalues $(\zeta_{k},  \bar{\zeta}_{k})$, where $\zeta_{k}  \in \mathbb{C}_{+}$  and $\bar{\zeta}_{k}=-\zeta_{k} \in \mathbb{C}_{-}$.  From conclusion (\ref{RTSymmetryEigenvetors}) in section 3, we  get the corresponding ``perturbed"  eigenvectors
\begin{eqnarray}\label{NRTSymmetryEigenvetors}
v_{k0}(\epsilon)=[1, e^{\sum_{j=0}^{N-1}b_{k,j}\epsilon^{j}} ]^{T},\  \bar{v}_{k0}=v_{k0}(-\bar{\epsilon}),\ \ b_{kj}\in \mathbb{C}.
\end{eqnarray}
Hence,  the $N$-th order $m$-solitons have $m(N+1)$  free complex constants,  $\{\zeta_{k},\ b_{k,j},\ 1\leq k\leq m,\  0 \leq j \leq N-1 \}$.

The  second-order fundamental-soliton is obtained when we set   $N=2, m=1$ with $n_{1}=\bar{n}_{1}=2$ in (\ref{highsoliton}), and the analytical expression is:
\begin{eqnarray}\label{RT2-ndHighSoliton}
&& q(x,t)=\frac{8\zeta_{1} e^{4i\zeta_{1}^2 t} \left[e^{-2i\zeta_{1}x-\ln b_{10}} f_{1}(x,t)+ e^{2i \zeta_{1}x+\ln b_{10}} \bar{f}_{1}(x,t) \right]}{4\cosh^2\left(2i\zeta_{1}x+\ln b_{10}\right)+ f_{0}(x,t)}.
\end{eqnarray}
where $f_{0}(x,t)=4(f_{1}(x,t)+i)(\bar{f}_{1}(x,t)+i)$, and
\begin{eqnarray*}
f_{1}(x,t)=\zeta_{1}(2x+8\zeta_{1}t-ib_{11}b_{10}^{-1})-i,\ \bar{f}_{1}(x,t)=\zeta_{1}(-2x+8\zeta_{1}t+ib_{11}b_{10}^{-1})-i.
\end{eqnarray*}
Although the fundamental-soliton in eq.(\ref{e:RTNLS}) are found to be stationary\cite{JYRTNLS2017}.  For this second-order fundamental-soliton, two solitons moving along the path
\begin{eqnarray}
 \Sigma_{\pm}:\ \  2  \textrm{Im} (\zeta_{1}) x  \pm \frac{1}{2}\ln  \left| f_{0}(x,t)\right|-\ln |b_{1}|=0  \label{RTAsmpLine1}
\end{eqnarray}
with almost the same velocity.  As $t\rightarrow \pm \infty$,  the  amplitudes $|q|$  changes as
\begin{eqnarray}
|q(x,t)|\sim \frac{8|\zeta_{1}|e^{-4\textrm{Im} (\eta_{1}^2)t}}{|e^{\pm 2i\gamma x-i\tau_{0} \pm 2i\arg(b_{10})}+1|},
\end{eqnarray}
where  $\gamma=2\textrm{Re}(\zeta_{1})$, $\tau_{0}= \text{Arg} \left[f_{0}(x,t)\right] +2 k \pi,\ (k \in \mathbb{Z})$.

This  soliton would also collapse at certain locations,  but not repeatedly collapse with time. Under  a suitable choice of parameters,  this high-order soliton can be non-collapsing. The amplitudes of two moving waves  grows or decays exponentially  when $\zeta_{1} \in \{\mathbb{C}_{+} \setminus i\mathbb{R}_{+}\}$, and it would decay/grow when $\zeta_{1}$ is in the first/second quadrant of the complex plane.   As concrete examples, graphs of these solitons are illustrated in Fig.6  with two sets of parameters.
\begin{figure}[htbp]
  \begin{center}
  \vspace{-2.25cm}
  \hspace{-7.0cm} \includegraphics[scale=0.260, bb=0 0 385 567]{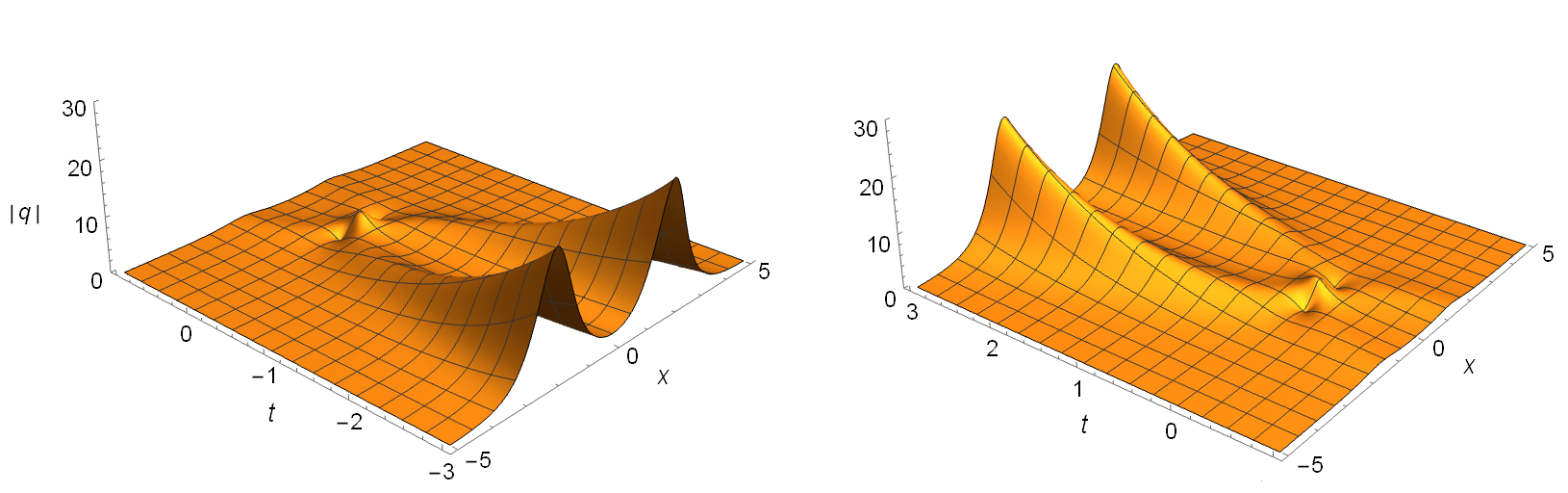}
   \caption{The  second-order one-soliton (\ref{RT2-ndHighSoliton}) with parameters:  (Left panel) $\zeta_{1}=0.1+i, e^{b_{10}}=1+0.1i,\ e^{b_{11}}=1$.
   (Right panel). $\zeta_{1}=-0.1+i, e^{b_{10}}=e^{b_{11}}=1$.  }
     \vspace{-0.75cm}
   \label{Tu6}
   \end{center}
\end{figure}

Normally,  the $N$-th order  fundamental-soliton could exhibit analogical features with the  second-order fundamental-soliton. There will be $N$ different asymptote trajectories  with $N$ waves moving along them in the nearly same velocities.  For instance, a decaying third-order one-soliton is displayed in Fig.7.
Moreover,  the high-order multi-solitons could exhibit  quite different dynamics.  For example,  the second-order two-solitons  move in four opposite directions when $\zeta_{1}$, $\zeta_{2}$ are not both purely imaginary,  and the repeated collapsing and ``four-way"  motion  can be observed.  Such an high-order  two-solitons solution is  shown in Fig.7,  which can not be seen  as a simple nonlinear superposition between two second-order fundamental-soliton.
\begin{figure}[htbp]
  \vspace{-1.75cm}
   \begin{center}
   \hspace{-3.55cm} \includegraphics[scale=0.275, bb=0 0 385 567]{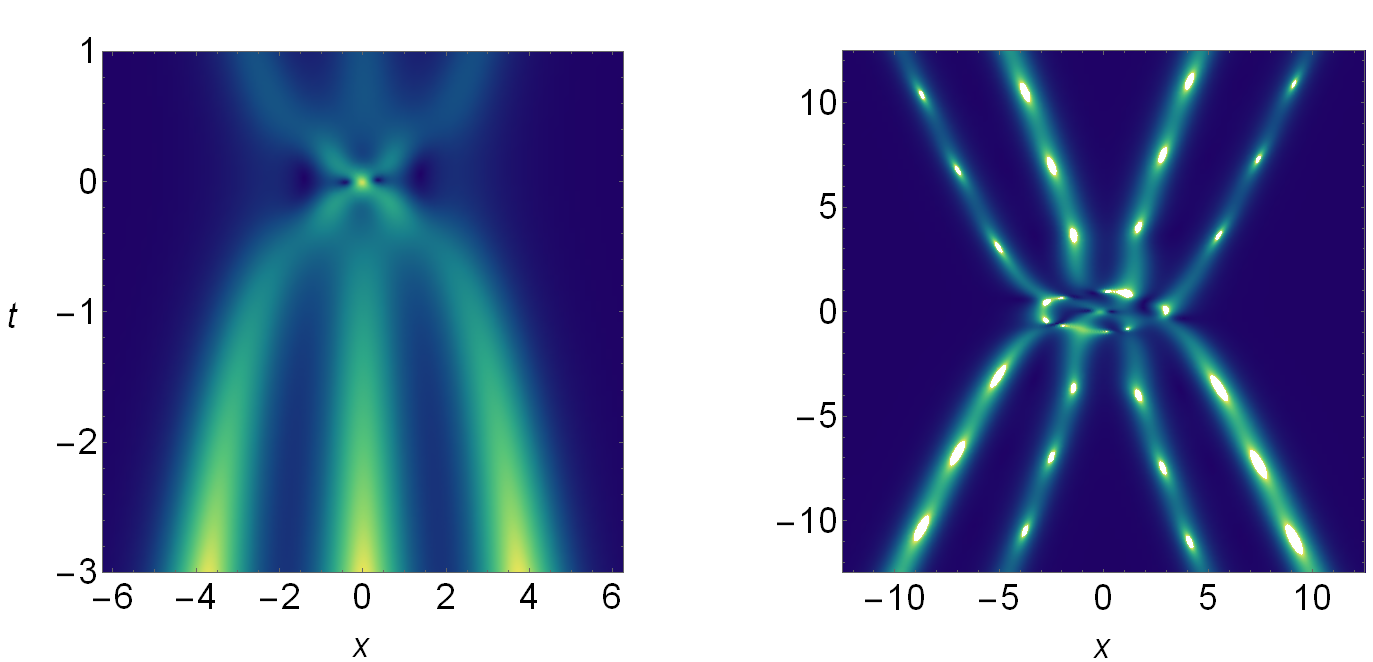}
   \caption{
   (Left panel). Density plot for the third-order soliton with parameters: $\zeta _1=0.05+i,\ e^{b_{10}}=1,\ e^{b_{11}}=0.1+0.1i.$
   (Right panel). Density plot for the second-order two-solitons   with parameters: $ \zeta _1=0.2+i,\ \zeta_2=-0.1+1.2i,\  e^{b_{10}}=1+0.5i,\ e^{b_{20}}=1,\ e^{b_{11}}=e^{b_{21}}=1$. }
     \vspace{-0.5cm}
   \label{Tu7}
   \end{center}
\end{figure}
\section{Dynamics of high-order solitons in the  reverse-space-time    NLS equation}
To derive  the $N$-th order solitons in the reverse-space-time  NLS equation (\ref{e:RSTNLS}), we  impose   symmetry  relations of discrete  scattering data (\ref{RSTSymmetryEigenvetors}) in the general soliton formula (\ref{highsoliton}).  In this case, the normal $N$-th order $m$-solitons have $2m$ free complex constants, $\{ \zeta_{k},\bar{\zeta_{k}}, 1\leq k \leq m \}$,  where $\zeta_{k}  \in \mathbb{C}_{+}$,  and $\bar{\zeta}_{k} \in \mathbb{C}_{-}$.

For the second-order fundamental soliton,  we choose $m=1$ with $N=n_{1}=\bar{n}_{1}=2$.  So the analytic expression for this solution is
\begin{eqnarray}\label{RST2-ndSoliton}
q(x,t)=\frac{4  \bar{\omega }_1 \left(\zeta _1-\bar{\zeta}_1\right) \left[\omega_1 \bar{\omega}_1 e^{-2 i \left(\bar{\zeta }_1x -2 \bar{\zeta }_1^2 t \right)}f_1(x,t)+ e^{-2 i \left(\zeta _1 x-2 \zeta _1^2 t\right)}\bar{f}_1(x,t)\right]}{e^{2 i \left(\bar{\zeta }_1-\zeta _1\right)x+4 i \left(\zeta _1^2-\bar{\zeta }_1^2\right)t}+ e^{ -2 i \left(\bar{\zeta }_1-\zeta _1\right)x-4 i \left(\zeta _1^2-\bar{\zeta }_1^2\right)t}+  \omega_1\bar{\omega}_1f_0(x,t)},\ \ \ \  \
\end{eqnarray}
where $f_{0}(x,t)=4(f_{1}+i)(\bar{f}_{1}+i)+2$,  and
\begin{eqnarray*}
f_{1}(x,t)=\left(\bar{\zeta }_1-\zeta _1\right)(x-4 \zeta_{1} t)-i,\  \bar{f}_{1}(x,t)=\left(\zeta _1-\bar{\zeta }_1\right)(x-4\bar{\zeta}_{1}t)-i,
\end{eqnarray*}
It is found that the above high-order fundamental-soliton (\ref{RST2-ndSoliton}) has two gradually paralleled center trajectories,  which  approximatively locate at following two curves:
\begin{eqnarray}
\Sigma_{\pm}:\ \ \textrm{Im}\left(\bar{\zeta }_1-\zeta _1\right)x -2\textrm{Im}\left(\bar{\zeta }_1^2  -\zeta _1^2\right)t \pm
\frac{1}{2}\ln\left[\left|f_{0}-2\right|\right]=0.  \label{RSTAsmpLine1}
\end{eqnarray}
Moreover,  regardless of the effect brought by the  logarithmic part as $t\rightarrow \pm \infty$,  two solitons  moving separately along $ \Sigma_{\pm}$ each curve in a nearly same velocity,  which is approximate to:
\begin{eqnarray*}
V \approx V_{c} :=  2\textrm{Im}\left(\bar{\zeta }_1^2  -\zeta _1^2\right)/\textrm{Im}\left(\bar{\zeta }_1-\zeta _1\right),
\end{eqnarray*}
and  the solution's amplitudes $|q|$  would approximately  changes as:
\begin{eqnarray}\label{Estimation|q|}
|q(t)|\sim 2|\zeta_{1}-\bar{\zeta}_{1}| \ \frac{e^{\beta t \pm \delta_{0}} }
{\left|e^{\pm 2i \gamma t-i \tau_{0}} + \omega_{1} \bar{\omega}_{1}\right|},    \ \ \  t \sim \pm \infty,
\end{eqnarray}
where,
\begin{eqnarray*}
&&\beta= -2V_{c} \textrm{Im}\left(\bar{\zeta}_1\right)-4\textrm{Im}\left(\bar{\zeta}_1^2\right), \ \gamma=V_{c}\textrm{Re}\left(\zeta_1-\bar{\zeta}_1\right)-2\textrm{Re}\left(\zeta_1^2-\bar{\zeta}_1^2\right). \\
&&\delta_{0}=-\textrm{Im}\left(\zeta_1+\bar{\zeta}_1\right)\ln \hspace{-0.1cm}\sqrt{|f_{0}-2|} /  \textrm{Im}\left(\bar{\zeta}_1-\zeta_1\right),
\end{eqnarray*}
with   $\tau_{0}=\textrm{Arg}\left[f_{0}(x,t)-2\right]+2k\pi,\ (k \in \mathbb{Z})$.

As can be seen from this estimation,   the amplitude of soliton is growing or decaying exponentially along $\Sigma_{\pm}$    at the rate of $e^{\beta t \pm \delta_{0}}$,  which depends mainly on the value of $\beta$ (except for $\textrm{Re}(\zeta_{1})=\textrm{Re} ( \bar\zeta_{1})$, where  $\beta=0$).  These are also some difference in the amplitudes when $q(x,t)$ moves  on different trajectories,  depending on the sign of $\delta_{0}$.  Especially, if $\delta_{0}=0$,  both of them  will keep the same amplitude.

Another interesting feature for this high-order fundamental soliton is the repeatedly collapsing phenomenon.  And the blowing-up interval $T_{c}$ for this solution admits  a ``perturbative" varying  period,  which can be roughly estimated as:  $T_{c} = \pi/|\gamma|+\Delta(t)$, where $\Delta(t)$  is a time-dependent small error term.  Regardless of minor changes in the arguments $\tau_{0}(t)$,  the approximately value of $\Delta(t)$  is attained  as $\Delta(t)\thickapprox \left[\bar{\tau}_{0}(t_{c}+\pi/|\gamma|)-\tau_{0}(t_{c})\right] / 2\gamma$,  where $t_{c}$  is the time coordinate for an initial singularity.  Examples are given for two sets of parameters:
\begin{eqnarray}
&& \zeta_{1} =-0.3+0.9i,\ \ \bar{\zeta_{1}} = -0.28-0.6i,\ \omega_1=\bar{\omega}_1=1, \label{RST-Parameters1} \\
&& \zeta_{1} =0.35+0.9i,\ \ \bar{\zeta_{1}} =0.325-0.6i,\ \omega_1=- \bar{\omega}_1=1. \label{RST-Parameters2}
\end{eqnarray}
Graphs of the two fundamental solitons are displayed respectively in  Fig.8.
\begin{figure}[htbp]
\begin{center}
   \vspace{1.0cm}
   \hspace{-6.0cm} \includegraphics[scale=0.1650, bb=0 0 385 567]{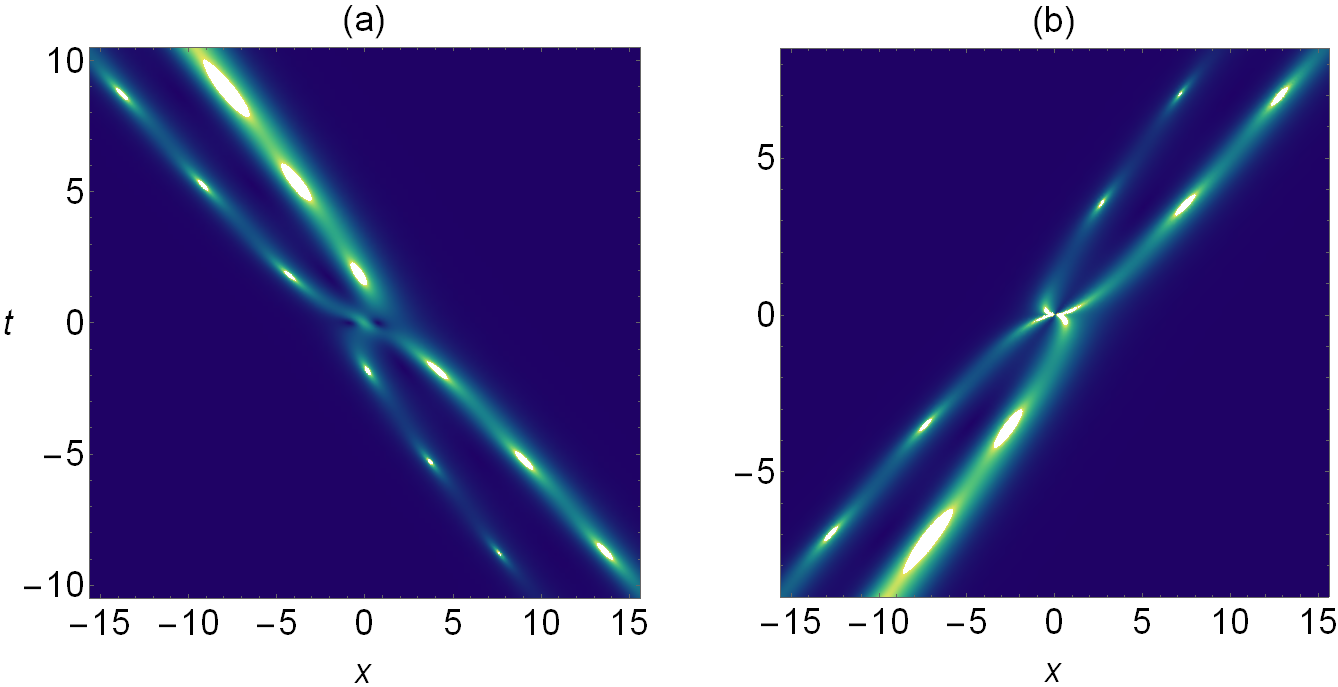}
   \caption{Two second-order one-solitons (\ref{RST2-ndSoliton}) in the reverse-space-time NLS equation (\ref{e:RSTNLS}).  The parameters for the density plot (a).  and (b).  are given by Eqs. (\ref{RST-Parameters1})  and (\ref{RST-Parameters2}) respectively.}
   \vspace{-0.75cm}
   \label{Tu8}
   \end{center}
\end{figure}
Apparently,    both of these two solitons collapse repeatedly with time. In the former solution, the soliton moves at velocity about $V_{c}\thickapprox -1.168$ (to the left).   The amplitude  $|q|$ exponentially increases along the curve $\Sigma_{\pm}$ at the rate of $e^{\beta t}$ with $\beta\thickapprox 0.0576$.  In the latter solution, the soliton moves at velocity $V_{c}\thickapprox 1.36 $ (to the right),  while $|q|$  decreases exponentially  along $\Sigma_{\pm}$ at the rate of $e^{\beta t}$ with $\beta\thickapprox -0.072$.

For the the high-order multi-solitons, because the eigenvalues   $\zeta_{k} \in \mathbb{C}_{+}$  and $\bar{\zeta}_{k} \in \mathbb{C}_{-}$  are totally independent, eigenvalues can be also arranged in several different configurations,  which give rise to new types of solitons for the reverse-space-time  NLS equation (\ref{e:RSTNLS}).  For instance, with symmetry (\ref{RSTSymmetryEigenvetors}) on the eigenvectors,  if we take $N=2$ with $I_{1}=\{1,1\}$ and $I_{2}=\{2\}$ in formula (\ref{highsoliton}),  certain choice of parameters can produce a high-order ``two-soliton".  Choosing $N=4$ with $I_{1}=I_{2}=\{2,2\}$,  we can  derive a nonlinear superposition between two different second-order one-soliton solutions(This solution can be also regarded  as a second-order  two-soliton).  Graphs of these solitons are very similar to those displayed in Fig.3  and Fig.4, so their novel dynamic behaviors can be expected.

\section{Summary and discussion}
In summary,  we have derived general high-order solitons in the \PT-symmetric, reverse-time, and reverse-space-time nonlocal NLS equations (\ref{e:PTNLS})-(\ref{e:RSTNLS}) by using a Riemann-Hilbert treatment.  We have shown that  through the symmetry relations on the ``perturbed" scattering data in each equation, the high-order solitons can be separately reduced from the Riemann-Hilbert solutions of the AKNS hierarchy.  At the same time, novel solution behaviours in these nonlocal equations have been  further discussed.
We have found that  the high-order fundamental-soliton  is always moving on several trajectories in nearly equal velocities,  while the high-order multi-solitons  could have  more complicated wave and trajectory structures.  In all these nonlocal equations,   a generic character in their high-order solitons is repeated collapsing.  Moreover, new types of high-order hybrid-pattern solitons are discovered, which can describe a nonlinear superposition between several types of solitons.  Our findings reveal the novel and rich  structures for high-order solitons  in the nonlocal NLS equations (\ref{e:PTNLS})-(\ref{e:RSTNLS}), and they could intrigue further investigations on solitons in the other nonlocal integrable equations.

In addition,  it should be noted that  by utilizing  new symmetry properties of scattering data in these nonlocal equations,
some open questions left over in previous Riemann-Hilbert derivations of solitons have been resolved in~\cite{JYRTNLS2017}.  That is,  when the numbers of eigenvalues (or, known as zeros  of the Riemann-Hilbert problem)  in the upper and lower complex planes, counting multiplicity,  are not equal to each other,   it would produce solutions which  are unbounded in space (thus never solitons).   Therefore,   in order to illustrate the validity for this conclusion in the case of multiple zeros,   we consider  the second-order fundamental-soliton in the \PT-symmetric NLS equation by choosing a single pair of eigenvalues $(\zeta_{1}, \bar{\zeta}_{1})\in i\mathbb{R}_{+}$ in   expression (\ref{RS2-ndHighSoliton}),  then  it produces a high-order ``fundamental-soliton".  Although it still satisfies eq.(\ref{e:PTNLS}),    this solution is not localized  in space and  grows exponentially in  the positive  $x$ directions.

\section*{Acknowledgment}
This project is supported by the Global Change Research Program of China (No.2015CB953904), National Natural Science Foundation of China
(No.11675054 and 11435005), and Shanghai Collaborative Innovation Center of Trustworthy Software for Internet of Things (No. ZF1213).

\end{document}